# STUDIES ON THE ANISOTROPIC PROPERTIES OF $MgB_2$

**O. F. de Lima**
Instituto de Física *Gleb Wataghin*
Universidade Estadual de Campinas - UNICAMP
13083-970 Campinas, SP, Brasil

## 1. INTRODUCTION

The discovery of superconductivity at 39 K, in Magnesium Diboride ($MgB_2$), [1] has brought new excitement to the area of basic and applied research on superconducting materials.

Several studies have already evidenced that $MgB_2$ has a good potential for applications [2,3,4,5] in view of the relatively high values of critical current density, $J_c$, and the successful preparation of wires and films. However, intense magnetic relaxation effects, associated with thermally activated flux creep and flux jumps, have been found to limit the $J_c$ values at high magnetic fields [6]. This means that effective pinning centers have to be added [7] into the material microstructure, in order to halt dissipative flux movements.

The observation of an isotope effect [8, 9], a BCS-type energy gap [10], as well as band structure studies [11,12,13], suggest the occurrence of a phonon-mediated superconductivity in $MgB_2$. Many other experimental [14,15,16,17,18,19] and theoretical [20,21] works have pointed also to the relevance of a phonon-mediated interaction, in the framework of the BCS theory. However, the possibility of hole superconductivity, associated with the nearly filled boron planar $p_{x,y}$ orbitals, has also been suggested [22]. Questions have been raised about the relevant phonon modes, as well as about the gap energy and Fermi surface anisotropies. A considerable amount of spectroscopic and thermal data has shown a broad range of values, between 2.5 and 5.0, for the gap ratio $2\Delta_0 / kT_c$, where $\Delta_0$ is the gap energy at $T = 0$ and $k$ is the Boltzmann constant. As a comparison, the BCS theory in the weak coupling limit predicts [23] $2\Delta_0 / kT_c \approx 3.5$, for an isotropic gap energy. In an effort to interpret this puzzling situation, multiple gap models [20,24] as well as a general model of anisotropic s-wave order parameter [18,21] have been proposed.

The strongly anisotropic crystalline structure of $MgB_2$ has been known for a long time. It seemed therefore reasonable when specific heat studies done in polycrystalline samples [15] as well as band structure calculations [11], pointed to the possible anisotropic nature of the electronic and magnetic properties of $MgB_2$. The first direct measurement of an anisotropic superconducting property was achieved for the bulk nucleation field $H_{c2}$, in samples of aligned $MgB_2$ crystallites [25]. It was found a ratio $H_{c2}^{ab} / H_{c2}^{c} \approx 1.7$, between the critical field parallel to the $ab$ plane and parallel to the $c$ axis direction. The anisotropic behavior of some normal state properties, like compressibility [26,27] and magnetoresistance [28], have also been reported.

Following, a brief review of some aspects regarding sample preparation will be presented in Section 2. In Section 3, a detailed analysis of the bulk nucleation field (or upper critical field) anisotropy will be described. In Section 4 other experimentally assessed anisotropic properties of $MgB_2$ will be briefly reviewed. Finally, in Section 5, a conclusion will be presented.

## 2. SAMPLE PREPARATION

The compound $MgB_2$ is known to occur in equilibrium with an excess of Mg, for temperatures above 650 °C [29]. By 1953 its $AlB_2$-type crystal structure was determined, using X-ray diffraction studies [30,31]. This layered structure belongs to the space group P6/mmm, and consists of alternating triangular layers of Mg atoms and *graphite-like* hexagonal layers of B atoms. The unit cell lattice parameters typically reported [30,31,32,33] are around $a = 3.085$ Å and $c = 3.521$ Å.

A common route to prepare $MgB_2$ polycrystalline samples starts by mixing the pure elements, Mg and B, in the atomic ratio B:Mg = 2:1. Because of the relatively high vapour pressure of Mg it is advised to react the mixture sealed in an evacuated tube or container of an appropriate material (e.g. Ta,

Nb, Fe, Mo, BN). In order to avoid oxidation of the tube/container external surface, especially in cases when open-air furnaces are used, the sealed tube/container is sealed inside a quartz ampoule, in a residual atmosphere of argon gas. Typically, a reaction time around 2 hours at temperatures between 700 °C and 1000 °C have been used, followed by cooling to room temperature. However, to obtain small single crystals, a higher reaction temperature between 1200 - 1400 °C and addition of Mg in excess to provide an internal vapour pressure above 1 bar, was reported in 1973 [32]. This is essentially the same route employed recently by different groups, to obtain $MgB_2$ crystallites having sub-millimeter sizes [25,33,34,35]. A different approach consists of reacting the quasiternary Mg-$MgB_2$-BN system, at pressures around 50 bar and temperatures ranging between 1400 - 1700 °C, for time periods between 5 to 60 minutes. In this case $MgB_2$ crystals reaching linear dimensions up to 0.7 mm has been reported [33].

$MgB_2$ thin films have also been grown successfully, using pulsed laser deposition [4,36] as well as by electron beam evaporation of B and Mg [37]. In some cases textured thin films were obtained, consisting of grains that have their crystallographic $c$ axis aligned perpendicularly to the substrate plane. An intriguing feature of thin films is that their $T_c$ ranges typically between 25 - 37 K, always below the bulk value of 39 K.

## 2.1. Aligned $MgB_2$ Crystallites

While polycrystalline $MgB_2$ is very easy to grow and is a readily available reagent, good-sized single crystals of this material, with linear dimensions above 0.7 mm have not yet been reported, and their development promises to keep being a real challenge. However, by the middle of February 2001, in the early days of intense activities on the new $MgB_2$ superconductor, samples of aligned $MgB_2$ crystallites were prepared at UNICAMP [25]. Firstly, a weakly sintered sample of $MgB_2$ was synthesised, starting from a stoichiometric mixture of 99.5 at% pure B and 99.8 at% pure Mg, both in chips form (Johnson Matthey Electronics). The loose mixture was sealed in a Ta tube under Ar atmosphere, which was then encapsulated in a quartz ampoule and put into the furnace. The compound formation was processed by initially holding the furnace temperature at 1200 °C for 1 hour, followed by a decrease to 700 °C (10 °C/h), then to 600 °C (2 °C /h), and finally to room temperature at a rate of 100 °C/h. The weakly sintered product was easily crushed and milled by employing mortar and pestle. Using a stereomicroscope one could observe at this stage a very uniform powder, consisting mainly of shiny crystallites with aspect ratios ranging from 2 to 5 and linear dimensions going up to 50 µm. The powder was then sieved into a range of particle sizes between 5 - 20 µm, which allowed the crystallites fraction to be maximised to almost 100%. Small amounts of the powder were then patiently spread on both sides of a small piece of paper, producing an almost perfect alignment of the crystallites laid down on top of the flat surface, as shown in the SEM picture of Fig. 1. The success of this method relies on the crystallites plate-like shape, which is indeed a macroscopic manifestation of its anisotropic crystalline structure. Fig. 2 shows an X-ray diffraction pattern ($q$ -2$q$ scan) measured on one of the aligned sample, displaying only the (001) and (002) reflections, coming from the $MgB_2$ phase. A lattice parameter c = 3.518 ± 0.008 Å was evaluated from these two peak positions. The two small impurity peaks marked with asterisks were indexed as $SiO_2$. The inset of Fig. 2 shows a rocking curve (ω scan) for the (002) peak that reveals an angular spread (Full Width at Half Maximum) around 4.6 degrees, associated with a small misalignment of the crystallites $c$ axis.

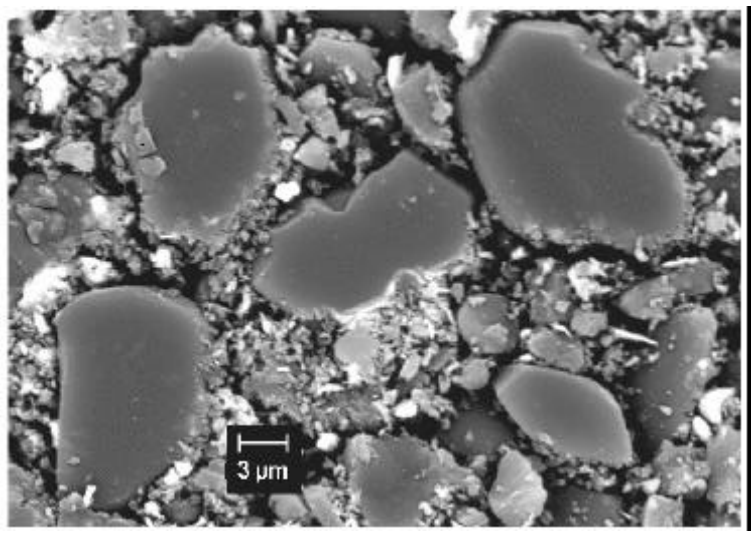

Fig. 1.   Scanning Electron Microscopy (SEM) picture showing the well-aligned crystallites and intercrystallite material.

Electron microprobe analysis done on four different areas between the $MgB_2$ crystallites, revealed the following average concentration (in at%) of elements: O (62.9), C (22.2), Ca (9.48), Si (1.48), Mg (1.44), Al (1.37), K (0.09), Fe (0.50), Cr (0.21), Ni (0.09). The first eight elements in this list were found also in the composition analysis made on the same type of paper used (Canson, ref. 4567-114). Microprobe analysis done also on the initial Mg and B revealed a few small precipitates, smaller than 10 μm and containing up to 8 at% Fe, only in the Mg chips. This confirms the expectation of Fe being a common impurity [38] in commercial Mg, and sets a general concern on its possible effects, although recent reports [5,39] have suggested no negative effects to the superconductivity of $MgB_2$ from Fe additions. The average composition found on top of several crystallites, normalised to the whole $MgB_2$ formula unit, was: Mg (30.80), O (2.20), Ca (0.17), Si (0.07), Fe (0.06). Although Boron contributes with a fraction of 66.6 at% it does not show-up in the microprobe analysis because it is too light. The contaminants found on top of the crystallites most possibly came from a surface contamination caused by the alignment technique, which required vigorous rubbing on top of the powder, using a steel tweezers tip to spread the crystallites uniformly. This is corroborated by a further analysis done on top of several as-grown crystallites, which detected only Mg and a small amount of O (possibly from MgO). This result is consistent with the very small solid solubility limit of about 0.004 at% Fe in Mg, which is known to occur [40] at the solidification temperature of 650 °C. The inter-crystallite type of *rubbish* shown in Fig. 1 is attributed mainly to the paper abrasion, which produces a varied distribution of irregular grains of paper fragments. In order to characterise the superconducting and magnetic properties of the aligned crystallites several samples were mounted, consisting of a pile of five squares of $3 \times 3$ mm$^2$, cut from the *crystallite-painted* paper and glued with Araldite resin. Each one of these samples contains a number of crystallites estimated to be around $6.5 \times 10^5$, ending up with an effective volume of 0.065 mm$^3$. This value is reasonably close to 0.060 mm$^3$ that was evaluated from the expected slope $M/H = -1/4\pi$, when the magnetization $M$ is measured at a very low field $H$.

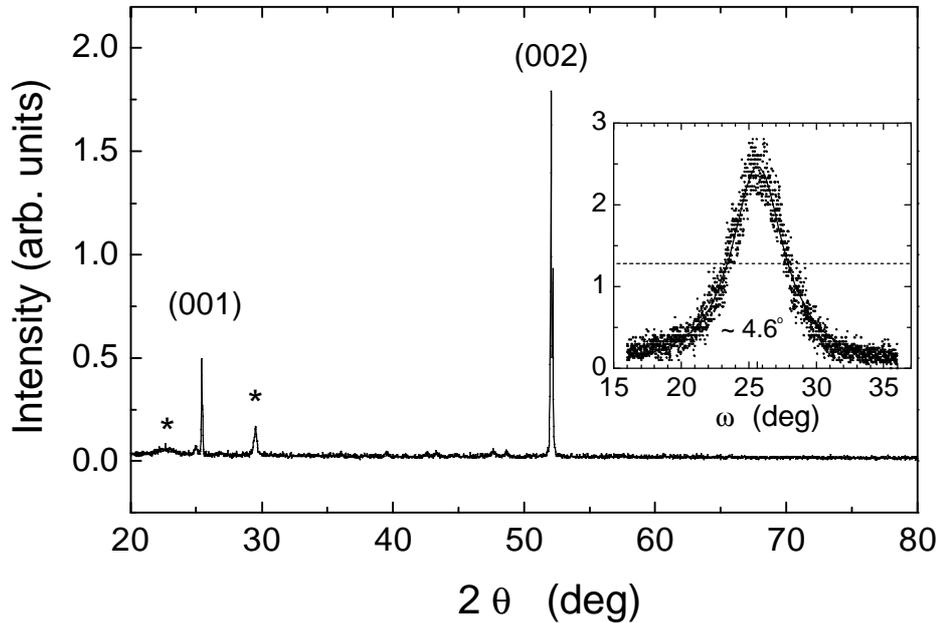

Fig. 2. X-ray diffraction pattern showing only the (001) and (002) peaks of $MgB_2$, plus two spurious peaks indexed as $SiO_2$.
Inset: rocking curve ($\omega$ - scan) for the (002) peak, showing an angular spread of about 4.6 degrees along the crystallites $c$ axis.

## 3. $H_{c2}$ ANISOTROPY

$H_{c2}$ anisotropy is one of the macroscopic manifestations of an anisotropic gap energy or an anisotropic Fermi surface, as well as a possible combination of both effects [41,42]. Therefore, its study is of paramount importance to help understanding the basic mechanisms involved in the pairing interaction. Samples of aligned MgB$_2$ crystallites, as described above, have been employed in detailed studies of the $H_{c2}$ angular dependence, providing a clear identification of its anisotropy factor and its bulk origin [25,43].

Fig. 3 shows the anisotropic signature of the $H_{c2}$ line in the field interval $0 \leq H \leq 40$ kOe. The experimental points were taken from the transition onset of the real component ($c'$) of AC susceptibility, measured using a PPMS-9T machine (Quantum Design), with an excitation field of amplitude 1 Oe and frequency 5 kHz. The inset shows an enlarged view of the $c'(T)$ curves for $H$ // $ab$ (open symbols) and $H$ // $c$ (solid symbols), the field orientation parallel to the $ab$ plane and parallel to the $c$ axis, respectively. The $c'(T)$ as well as the $M(T)$ (inset of Fig. 4) measurements, for $H = 10$ Oe, show sharp transitions at the same critical temperature $T_c \approx 39$ K. Typically, some of the published data on the temperature dependence of $H_{c2}$ [44,45] agree with the result displayed in Fig. 3 for $H_{c2}$ // $ab$. As an example, the data from Ref.[45] is plotted in Fig. 3 as stars. This mean that in polycrystalline samples the transitions are broadened, showing the onset at the highest temperature that corresponds to the highest critical field available, which is $H_{c2}$ // $ab$.

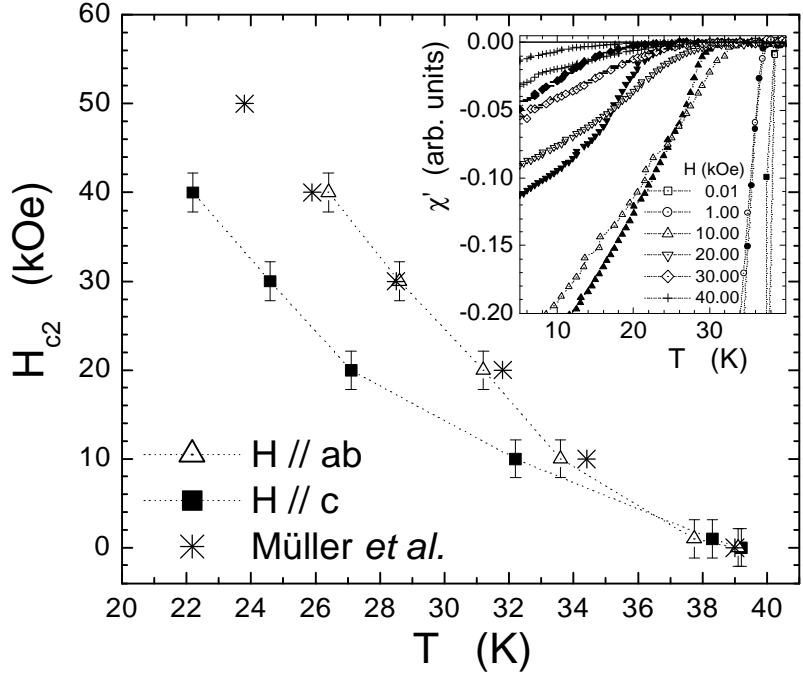

Fig. 3. Upper critical field $H_{c2}$ vs. Temperature phase diagram, for both sample orientations. The stars represent the $H_{c2}$ vs. $T$ line from Ref. [45]. The inset shows the real component $c'$ of the ac susceptibility vs. temperature, measured at several DC fields for both orientations. Open symbols are for the $H // ab$ curves and solid symbols for $H // c$.

The ratio $H_{c2}^{ab} / H_{c2}^{c}$, between the upper critical field when $H$ is applied parallel to the $ab$ plane and when it is along the $c$ direction, was evaluated at different temperatures, producing a value around 1.7. From [23] $H_{c2}^{ab} / H_{c2}^{c} = x_{ab} / x_c$, $H_{c2}^{ab}(T) = \Phi_0 /(2p\, x_{ab}\, x_c)$, $H_{c2}^{c}(T) = \Phi_0 /(2p\, x_{ab}^2)$, and using the Ginzburg-Landau (G-L) mean field expression for the coherence length, $x(T) = x_0 (1 - T / T_c)^{-0.5}$, one finds $x_{0,ab} \approx 70$ Å and $x_{0,c} \approx 40$ Å, the coherence length at $T = 0$ in the $ab$ planes and along the $c$ axis, respectively. The quantum of flux, in CGS units, is $\Phi_0 = 2.07 \times 10^{-7}$ G cm$^2$. A mass anisotropy $e^2 = (H_{c2}^{c} / H_{c2}^{ab})^2 \approx 0.3$ is then found for MgB$_2$, which could be considered a mild anisotropy when compared to the highly anisotropic high-$Tc$ cuprates [46], like YBa$_2$Cu$_3$O$_{7-\delta}$ ($e^2 \approx 0.04$) and Bi$_2$Sr$_2$CaCu$_2$O$_{8+\delta}$ ($e^2 \approx 10^{-4}$).

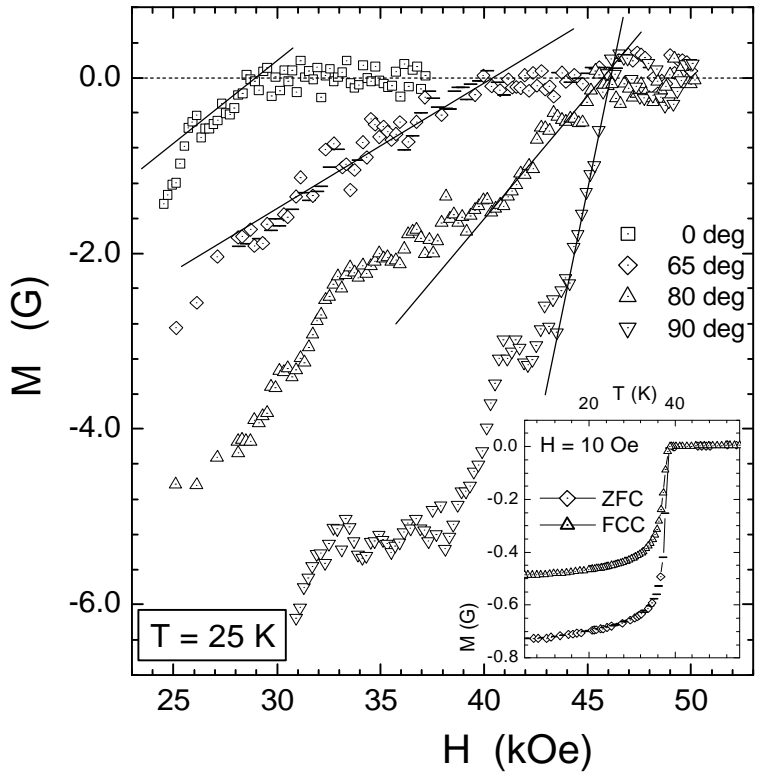

Fig. 4. Zero Field Cooling magnetization measurements as a function of the applied field, for $q$ = 0, 65, 80, 90 degrees. The bulk nucleation field $H_{c2}(q)$ is defined at the crossing of the auxiliary straight lines and the horizontal baseline ($M = 0$). The inset shows ZFC and FCC magnetization measurements as a function of temperature for $H = 10$ Oe, giving $T_c \approx 39$ K.

In order to study the $H_{c2}$ angular dependence, under axial applied fields, a special sample rotator was built [43] with all parts machined from a low magnetic teflon rod. Magnetization measurements were performed with a SQUID magnetometer (model MPMS-5, made by Quantum Design). Fig. 4 shows the magnetic dependence of the magnetization in $T = 25$ K, for a few representative angles, $q$, between the sample $c$ axis and the magnetic field direction. The ZFC (Zero Field Cooling) measurements shown in the main frame look noisy possibly due to the effect of intense vortex creep [6], combined with a complex regime of flux penetration in the granular sample of aligned crystallites. Thus, the occurrence of random weak links and the varied coupling between grains contribute also to produce a fluctuating behavior in the sample overall response. However, in all cases it was possible to define $H_{c2}(\boldsymbol{q})$, at the crossing point between the horizontal baseline and the straight line drawn across the experimental points in the region near the onset of transition. This linear behavior of the magnetization close to the onset is indeed expected from the G-L theory [23]. A constant paramagnetic background was subtracted from all sets of data. In fact one of the reasons for measuring at T = 25 K is because at this temperature $H_{c2}(\boldsymbol{q})$ ranges between 28 - 36 kOe, where the paramagnetic background is already saturated [25].

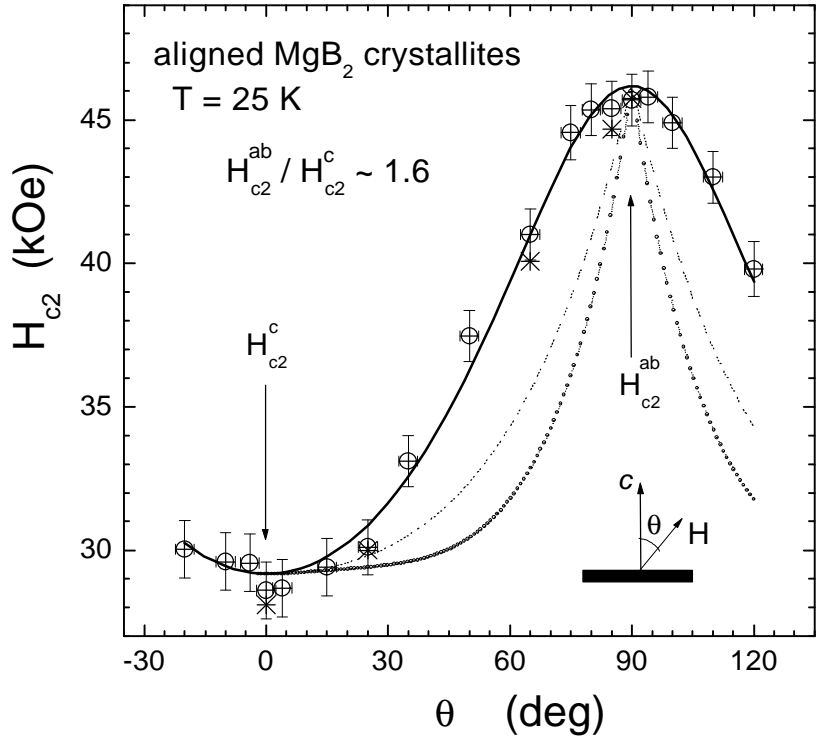

Fig. 5. Bulk nucleation field (or upper critical field), $H_{c2}$, as a function of the angle, $\boldsymbol{q}$, between the sample $c$ axis and the magnetic field direction. Plots of the expected angular dependence for the surface nucleation field, $H_{c3}$, in thick samples (dash-dotted curve) and very thin films (dashed curve) are also shown. The stars at $\boldsymbol{q} = 0$, 25, 65, 85, 90 degrees represent $H_{c2}(\boldsymbol{q})$ obtained at the onset of transition of the real part of $ac$ susceptibility measurements.

Fig. 5 displays $H_{c2}(\boldsymbol{q})$ for $\boldsymbol{q}$ between -20 deg and 120 deg. The vertical error bars were estimated to be around ±1 kOe while the horizontal error bars, of ±2.5 deg, almost coincide with the symbol size. The solid line going through the experimental points represents a good fit of the angular dependence, predicted by the 3D anisotropic G-L theory to be [23,41]

$$H_{c2}(\boldsymbol{q}) = H_{c2}^{c} \left[ \cos^2(\boldsymbol{q}) + \boldsymbol{e}^2 \sin^2(\boldsymbol{q}) \right]^{-1/2}, \qquad (1)$$

where $\boldsymbol{e}^2$, is the mass anisotropy ratio. The best fitting is obtained with $\boldsymbol{e}^2 \approx 0.39$, implying that $H_{c2}^{ab} / H_{c2}^{c} \approx 1.62$, which is close to the value 1.7 anticipated by AC susceptibility measurements done for the two extreme $\boldsymbol{q}$ positions, at 0 and 90 degrees (see Fig. 3). In Fig. 5 we see also five data points ($\boldsymbol{q}$ = 0, 25, 65, 85, 90 deg) marked with stars, which were obtained at the onset of transition of the real part of the complex susceptibility, measured with an excitation field of amplitude 1 Oe and frequency 5 kHz.

The ratio $H_{c2}^{ab} / H_{c2}^{c} \approx 1.62$ reminds the relationship predicted for the surface nucleation field [47] $H_{c3} \approx 1.7 H_{c2}$. However, this is clearly not the case for the presented data, as one can see from the dash-dotted curve in Fig. 5, which is a plot of the expected angular dependence of the surface nucleation field for thick samples, given by [48]:

$$\left[ \frac{H(\boldsymbol{q})}{H_{c3}} \sin \boldsymbol{q} \right]^2 \left[ 1 + \cot \boldsymbol{q} \left( 1 - \cos \boldsymbol{q} \right) \right] + \frac{H(\boldsymbol{q})}{H_{c2}} \cos \boldsymbol{q} = 1 \ , \qquad (2)$$

where $H_{c3} = H(\boldsymbol{q} = 90°)$ and $H_{c2} = H(\boldsymbol{q} = 0°)$. The dashed curve in between represents the well-known Thinkham's formula[23,49]

$$\left[ \frac{H(\boldsymbol{q})}{H_{c}^{ab}} \sin \boldsymbol{q} \right]^2 + \left| \frac{H(\boldsymbol{q})}{H_{c}^{c}} \cos \boldsymbol{q} \right| = 1 \ , \qquad (3)$$

which is valid for the surface nucleation field in very thin films. From both plots one sees that a characteristic feature of the surface nucleation field is a cusplike curve shape near $\boldsymbol{q}$ = 90 degrees. This behavior contrasts with the sinusoidal shape followed by the MgB$_2$ data displayed in Fig. 5. Thus, a strong support is given to interpret the observed upper critical field as a genuine bulk nucleation field.

Several other studies have already confirmed the $H_{c2}$ anisotropy in MgB$_2$. Measurements of resistive transitions on three c-axis oriented thin films [50] showed anisotropy ratios ($H_{c2}^{ab}/H_{c2}^c$) of 1.8, 1.9 and 2.0, the ratio increasing with higher resistivity, correlated with a more impure or alloyed sample. Resistivity measurements done on single crystals, having the major linear size around 0.5 mm [33,34], showed anisotropy ratios close to 2.6. In Ref.[34] the authors have also shown anisotropic $H_{c1}$ lines pointing to $H_{c1}^{ab}/H_{c1}^c \approx 1.4$, where $H_{c1}$ is the lower critical field. However, one should be aware about the large uncertainty usually associated with $H_{c1}$ evaluations, mainly due to two factors: the very slow departure from linearity, when leaving the Meissner regime, and the strong dependence on demagnetizing fields. Measurements of resistive transitions on very clean epitaxial thin films [28], gave a relatively smaller anisotropy ratio around 1.3 in a large temperature region between 2 - 32 K. A much smaller anisotropy ratio value around 1.1 was also reported on a partially textured, hot deformed, bulk material [51].

The relatively large scattering of values for the $H_{c2}$ anisotropy ratio, going from 1.3 to 2.6, could possibly be ascribed to at least three factors. One is the sample purity, since it affects directly the energy gap anisotropy at the microscopic level [41,42]. The second is the experimental criterion used to define $H_{c2}$ or $T_{c2}$, such that a reliable bulk transition point is actually guaranteed. The third factor is a possible temperature dependence of the anisotropy ratio that could be originated from a temperature dependent gap anisotropy [21,52]. In this case, results obtained with samples of different purity levels and measured at different temperatures could not be directly compared.

The macroscopic $H_{c2}$ anisotropy can be caused by an anisotropic gap energy or by an anisotropic Fermi surface, as well as by a combination of both effects. Assuming an isotropic gap, one gets $x_{ab}/x_c = V_F^{ab}/V_F^c$, since [23] $x \propto V_F/\Delta_0$. Therefore, the data for aligned MgB$_2$ crystallites [25,43] implies $V_F^{ab} \approx 1.6 V_F^c$, for the Fermi velocities within the $ab$ plane and along the $c$ direction. However, several experimental [15,17,18] and theoretical [11,12,21] works have suggested an anisotropic gap energy for MgB$_2$. In particular, two recent reports [18,21] rely on the analysis of spectroscopic and thermodynamic data to propose an anisotropic s-wave pairing symmetry, such that the minimum gap value, $\Delta_0 \approx 1.2 kT_c$, occurs within the $ab$ plane. Using this result and assuming an isotropic Fermi surface the expected $H_{c2}$ anisotropy would be $H_{c2}^{ab}/H_{c2}^{c} \approx 0.8$. This conflicts with all experimental results that show clearly $H_{c2}^{ab} > H_{c2}^{c}$. However, by allowing a Fermi surface anisotropy in their model, Haas and Maki have found that [52] $V_{F}^{ab} \approx 2.5\,V_{F}^{c}$ in order to match the ratio $H_{c2}^{ab}/H_{c2}^{c} \approx 1.6$, at $T = 25$ K. Therefore, it seems that the two fundamental sources of microscopic anisotropy is affecting the $H_{c2}$ anisotropy of MgB$_2$ in opposite ways. As a consequence of combining both effects to explain the $H_{c2}$ anisotropy, the Fermi velocity anisotropy becomes about 60% higher when compared with the isotropic gap hypothesis. Interestingly, a calculation based on a two-band model has also found [53] $V_{F}^{ab} \approx 2.5\,V_{F}^{c}$, while a much smaller value of $V_{F}^{ab} \approx 1.03\,V_{F}^{c}$ was found in a band structure calculation using a general potential method [11].

## 4. OTHER ANISOTROPIC PROPERTIES

Another anisotropic property already assessed in the superconducting state of MgB$_2$ is the field penetration depth $l$ [54]. Using a radio frequency technique $l(T)$ was measured in polycrystalline samples and its anisotropic values were evaluated, from a theoretical analysis, to be around $l_{ab} = 1200\,\text{Å}$ and $l_c = 2450\,\text{Å}$.

Besides the strongly anisotropic crystalline structure of MgB$_2$ three other normal state anisotropic properties have been identified, the magnetoresistance, the compressibility and the thermal expansion. The magnetoresistance measured at $T = 45$ K, in epitaxial thin films, was found [28] to increase monotonically up to 8% for $H // c$ and up to 13% for $H // ab$, when H goes up to 60 T. Concerning the compressibility studies, similar results were obtained by two groups, one [26] using synchrotron X-ray diffraction and applying up to 6 GPa of pressure on the sample and the other [27] using neutron powder diffraction and applying up to 0.6 GPa of pressure. They followed the relative lattice parameters variation as a function of applied pressure, at room temperature, and found that the compression along the $c$ axis is about 64% larger than along the $a$ axis. Ref.[27] reports also that the thermal expansion (between 11 K - 297 K) along the $c$ axis is about twice that along the $a$ axis.

Although no experimental results on $J_c$ anisotropy was reported yet, it could be anticipated that the in-plane critical current density values are expected to be at least about 60% higher than the values along the $c$ axis direction ($H // ab$). This result is expected because $J_c$ is proportional to $x^2$, therefore [46] $J_c(H /\!/ c)/J_c(H /\!/ ab) \approx x_{ab}/x_c \approx H_{c2}^{ab}/H_{c2}^{c}$. This means that in order to optimise $J_c$ in wires or other polycrystalline components some texturization technique will be useful.

## 5. CONCLUSION

During the last four months an intense research activity has been devoted to the binary compound $MgB_2$, since it was found to be a superconductor for temperatures below 39 K. The anisotropic crystalline structure of $MgB_2$ consists of triangular layers of Magnesium atoms sandwiched between hexagonal layers of Boron atoms. Therefore, it should indeed be expected some anisotropy in its physical and chemical properties.

This paper presented a brief review on the already reported anisotropic properties of $MgB_2$, in the superconducting state (e.g. upper critical field $H_{c2}$, field penetration depth $l$, coherence length $x$, and energy gap $D$) as well as in the normal state (e.g., magnetoresistance, compressibility, thermal expansion). So far (as by May, 2001), the reported results have been carried out using aligned crystallites, c-axis oriented thin films and sub-millimiter crystals. However, good-sized single crystals with linear dimensions above 1 mm are still highly desirable. This will open the possibility of new advances, making easier to probe directly other expected anisotropic properties, like the critical current density and the normal state resistivity, among others. Since impurity scattering affects directly the gap anisotropy it will be also of interest to know the effects of crystal purity on the anisotropy factors. Most possibly this could be one of the main reasons for the scattered values observed in the $H_{c2}$ anisotropy ratio, ranging between 1.3 and 2.6.

## ACKNOWLEDGEMENTS

I would like to thank my close collaborators C. A. Cardoso, R. A. Ribeiro, M. A. Avila and A. A. Coelho, for their essential contributions in our anisotropy studies on the new $MgB_2$ superconductor. My thanks also to S. Haas, A. V. Narlikar, E. Miranda and O. P. Ferreira, for many useful discussions. I acknowledge the financial support from the Brazilian Science Agencies FAPESP and CNPq.